# Teaching Physics During the 17th Century: Some Examples From the Works of Evangelista Torricelli

**Amelia Carolina Sparavigna**
Department of Applied Science and Technology, Politecnico di Torino, Italy

**Abstract**: This paper is proposing some physics problems from a book, De Motu Gravium Naturaliter Descendentium, written by Evangelista Torricelli, physicist and mathematician best known for his invention of barometer, with the aim of showing how physics was taught in the Italian universities, those of Padua and Pisa for instance, during the 17th Century.

1. Introduction

How was the teaching of physics at the universities of Padua or Pisa, those where Galileo Galilei taught, during the 17th Century? To answer this question, which is anything but simple, let us try some comparisons with the lectures that we are giving today at university.
To have a proper picture of the scholar environment at that time, let us first consider that during the 17th Century physics did not use derivatives, integrals and plots of functions, and that the theory was based on geometric rules and proportions and on the results of some experiments and observations of natural phenomena. In the books, the subjects were discussed in Latin, which was the scientific language in Europe.
Besides a discussion in the framework of history of science, in my opinion it is possible to answer the question proposed above by using some practical examples too. We can show, for instance, some problems concerning the motion of masses on inclined planes, as they were taught by the teachers of that period to their students. To give these examples we have at our disposal a book written by Evangelista Torricelli, entitled De Motu Gravium Naturaliter Descendentium, which is quite suitable for this purpose. This text is available in the Torricelli's Opera Geometrica, published among the Google eBooks.

2. Evangelista Torricelli

Torricelli was born in 1608 in Faenza, at that time under the Papal States. The family was very poor, but seeing the child's talents, he was sent to study under the care of an uncle, who was a Camaldolese monk. From 1624 to 1626, he studied mathematics and philosophy in a college, and then the uncle sent him to Rome to study science under Benedetto Castelli (1578–1643), professor of mathematics at the Collegio della Sapienza. In 1641, Castelli asked Galileo to receive Torricelli, and for this reason, Torricelli moved to Florence, where he met Galileo, and was his secretary during the last months of Galileo's life.
After Galileo's death in 1642, the Grand Duke of Tuscany, Ferdinando II de' Medici, asked Torricelli to succeed Galileo as professor at the first occasion, and hired him as ducal mathematician. During this office, besides solving some mathematical problems, he also worked on telescopes and simple microscopes [1,2].

The Galileo's ideas and methods strongly influenced Torricelli's physics, as we can see from his studies collected in the Opera Geometrica, 1644. This volume contains the works entitled De Sphaera et Solidis Sphaeralibus, De Motu Gravium Naturaliter Descendentium, et Projectorum, De Dimensione Parabolae, and De Solido Hyperbolico cum Appendicibus de Cycloide, & Cochlea.

Torricelli is famous because he invented the mercury barometer in 1643, and is honored with a unit of measure for the pressure, the torr, named after him. He was also a pioneer in the area of infinite series.

Torricelli was a strong supporter of the geometry of indivisibles, developed by Bonaventura Cavalieri [3,4], and he applied this geometry to the determination of the center of gravity of some bodies. In a letter to Michelangelo Ricci of 1646, he communicated that, using the indivisibles, it is possible to determine a "universal theorem," which allows finding the center of gravity of any figure. Among the cases he mentioned, there is that of the circular sector [5].

### 3. Impetus, momentum and impact force

Other Torricelli's works were collected and edited by Tommaso Bonaventuri in a volume, in the year 1715 [6]. The preferred subject of these works was physics. The volume includes eight lectures proposed during some sessions of the Accademia della Crusca [7], of which he was a member. Among these lectures, there were three of them on the force of impact, two on lightness and one on the wind. In the lectures on the impact force (forza della percussione), Torricelli is also reporting some conversations he had with Galileo Galilei. In fact, several times in these lectures and in the Opera Geometrica, we find Torricelli writing that what he is discussing he had learned from Galileo, in particular about the motion of freely falling bodies, the parabolic motion of projectiles and the motion on inclined planes.

As previously told, Torricelli appreciated the geometry of indivisibles [4], which were the precursors of Leibniz's integration and Newton's derivation. However, as we can see from the lectures in [6], he did not apply indivisibles to physics.

Let us consider for instance, what Torricelli is telling about the impact force. Torricelli tries to analyze impetus [8] and momentum of a freely falling body, starting from rest at a certain height. He tells that impetus and momentum are increasing during the fall, because the gravity is continuously acting on the body. For this reason, their variation is infinite, as deduced by Galileo. However, the impact of the falling body on another body is producing a finite effect. Torricelli solved this problem of an infinite increase of the momentum with a finite effect during impact, arguing that the increase is given as infinite, because the speed of the body is passing from an initial null value to a final finite value. About impact, Torricelli tells that its effect could be infinite, only in the case the time of collision were null, but this is impossible.

Actually, Torricelli was not able to use the geometry of indivisibles to calculate the change of momentum as a sum of elemental contributions, to obtain a finite value of its variation, instead of an infinite one. In any case, he argued that the effect of impact is larger according to the height the body is falling.

Torricelli is also studying the role of time during the impact, and then he was, in a certain manner, distinguishing the force from the impulse, which is the integral of force over time of impact. From his lectures then, we can argue that Torricelli's concept of impetus was closer to the modern concept of impulse than that of force.

The lectures we find in Ref.6 were given to the member of the Accademia della Crusca. However, we are aiming to imagine a possible lecture of physics to students of the universities of that time. For this reason, the lectures to the Accademia are not enough. We need to use the Torricelli's books too, to extrapolate some practical discussions with diagrams and calculations that he could had given to his students. Let us remark once more that Torricelli had not the differential calculus: he used geometry and proportions.

In the following, it is proposed to the reader a practical approach, that is, the discussion of some problems prepared after Torricelli's book entitled De Motu Gravium Naturaliter Descendentium [9]. Here the aim is not a complete translation or analysis of this book, to which a future work is devoted, but the comparison of Torricelli's discussions with our discussions of mechanics. Some images are reproduced from [9]. Among the subjects, we will find the motion on inclined planes, the center of mass and equilibrium of forces.

In the following, let us consider that the Torricelli's momentum changes are linked to accelerations, as we can argue from his lectures to the Accademia della Crusca.

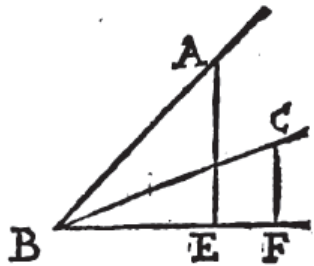

**Fig.1 – Two inclines planes having the same length but different angles.**

### 4. Problem on Momenta

The Latin text of the problem is the following. "Sint plana *ab*,*cb* inaequaliter inclinata, & sumptis aequalibus *ab*,*cd* ducantur perpendicula *ae*, *cf*: ad horizontem *bf*. Supponit Galileus pro demonstrata, momentum in plano *ab* ad momentum in plano *cb* ita esse ut est *ae* ad *cf*."

Let us imagine two inclined planes having equal lengths (Fig.1), but different angles of inclination. Being their height *ae*,*cf*. Galileo supposed that the ratio of the momenta on the plane *ab* and on *cd* is the ratio of *ae* and *cf*.

Torricelli assumes this as demonstrated by Galileo. However, we can use this example to understand what Torricelli's momentum was. Let us then try to show that the ratio of momenta is the same as the ratio of the heights of inclined planes, using accelerations.

Let us consider the momentum as $m \cdot \Delta v$ the product of mass and the increase of velocity during the time interval $\Delta t$. The increase of velocity is due to acceleration ($\theta = \alpha, \beta$ are the angles of the inclined planes, *g* gravitational acceleration).

First, let us consider a point-like mass, and no friction:

$$m \cdot \Delta v_{AB} = mg \sin \alpha \cdot \Delta t$$
$$m \cdot \Delta v_{CB} = mg \sin \beta \cdot \Delta t$$
(1)

Then:

$$\frac{m \cdot \Delta v_{AB}}{m \cdot \Delta v_{CB}} = \frac{mg \sin \alpha \cdot \Delta t}{mg \sin \beta \cdot \Delta t}$$
$$= \frac{\sin \alpha}{\sin \beta} = \frac{AE}{AB}\frac{BC}{CF} = \frac{ae}{cf}$$
(2)

In fact, the lengths are the same: *AB=BC*.

Of course, we can imagine that Galileo made some experiments to test the result. Therefore, he could had used some rolling spheres on the inclined planes. The acceleration of the center of the sphere is $a = \kappa g \sin\theta$, where $\kappa = (1+k^2)^{-1}$ and $k^2 = 2/5$. Then again, considering the momentum of the center of the sphere:

$$\frac{m \cdot \Delta v_{AB}}{m \cdot \Delta v_{CB}} = \frac{\kappa \cdot mg \sin\alpha \cdot \Delta t}{\kappa \cdot mg \sin\beta \cdot \Delta t}$$
$$= \frac{\sin\alpha}{\sin\beta} = \frac{AE}{AB}\frac{BC}{CF} = \frac{ae}{cf} \qquad (3)$$

### 5. Problem on two bodies connected by a massless string

"Duo gravia simul coniuncta ex se moveri non posse, nisi centrum commune gravitatis ipsorum descenda. Quando enim duo gravia ita inter se coniuncta fuerint, ut ad motum unius motus etiam alterius consequatur, erunt duo illa gravia tamquam grave unum ex duobus compositum, sive id libra fiat, sive trochleam, sive qualibet alia mechanica ratione, grave autem huiusmodi non movebitur unquam, nisi centrum gravitatis ipsius descendat. Quando vero ita constitutum fuerit ut nullo modo commune ipsius centrum gravitatis descendere possit, grave penitus in sua positione quiescet: alias enim frustra moveretur; horizontali, scilicte latione, qua nequaquam deorsum sendit. Propositio I: Si in planis inaequaliter inclinatis, eandem tamen elevationem habentibus, duo gravia constituantur, quae inter se eandem homologe rationem habeant quam habent longitudines planorum, gravia aequale momentum habebunt."

Let us consider two inclined plane with different angles, but the same height. On them we have two masses, linked by a massless rope, in such a manner that $m_a g/(m_b g) = a/b$, where *a,b* are the lengths of the plane. The two masses are in equilibrium.

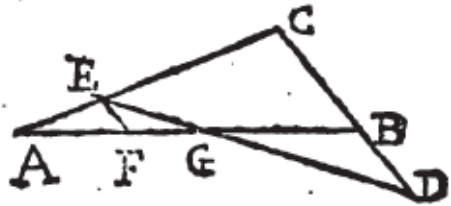

**Fig. 2 – The geometry of the problem.**

"Sit *ab,* horizon, & plana inaequaliter inclinata *ca,cb*. Fiat ut *ac* ad *cb*, ita grave aliquod *a*. ad grave *b*. Et gravia haec in homologis planis collocentur, in punctis a, & b, eiusdem horizontalis linae (Fig.2). Connectantur etiam aliquo immaginario funiculo per *acb*. ducto, adeo ut ad motum unius motus alterius consequatur. Dico gravia sic disposita aequale momentum habere: hoc est in ea in qua sunt positione aequilibrata conquiescere, neq sursum aut deosrum moveri. Ostendemus enim centrum commune gravitatis eorum descendere non posse, sed in eadem semper horizontali linea (quantumlibet gravia moveantur) reperiri."

Torricelli tells that the masses are in equilibrium because they have the same "momentum".

Moreover, if we move the masses on the planes, the height of the center of mass of this system does not change.

Let us consider the following notation: $m_a, m_b$ where $m_a g/(m_b g) = a/b$ (see the following Fig.3 too).

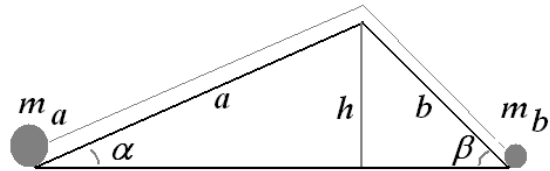

**Fig.3 – In this problem, we have two masses connected by a rope.**

As we did before, let us consider their momenta as the components of weights parallel to the inclined planes, multiplied by $\Delta t$ :

$$
\begin{aligned}
momentum_a &= (m_a \cdot \Delta v_a) \\
&= m_a g \sin\alpha \cdot \Delta t \\
momentum_b &= (m_b \cdot \Delta v_b) \\
&= m_b g \sin\beta \cdot \Delta t
\end{aligned}
\quad (4)
$$

For the equilibrium, we must have that $momentum_a = momentum_b$. Then:

$$
m_a g \sin\alpha = m_b g \sin\beta
$$
$$
m_a g /(m_b g) = \sin\beta / \sin\alpha = \frac{h}{b}\frac{a}{h} = a/b. \quad (5)
$$

The first equation means that the two masses have an equal momentum when the ratio of masses (or weights) is equal to the ratio of the lengths of inclined planes having the same height *h*, as given by Torricelli. This equation is also the one we obtain considering the components of weights parallel to the inclined planes, that is, the equilibrium of forces.

Let us choose a different position for the two masses at rest, for instance, moving upwards of a length *s* on the inclined plane the mass $m_a$, the other mass $m_b$ downwards of the same quantity *s*, because they are linked by the rope (Fig.4). Is the center of mass moving vertically? No.

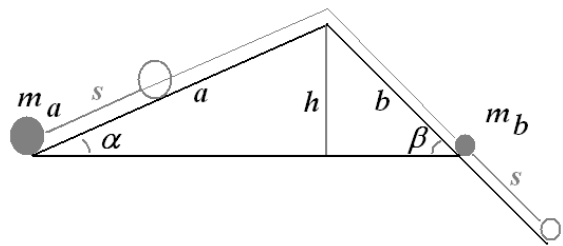

**Fig.4 – Let us move the masses in a different position.**

Let us call *y* a vertical axis. Then, the center of mass has a height:

$$\begin{aligned} y_{cm} &= \frac{m_a \sin\alpha \cdot s - m_b \sin\beta \cdot s}{m_a + m_b} \\ &= \frac{s}{m_a + m_b}(m_a \sin\alpha - m_b \sin\beta) \\ &= \frac{s}{m_a + m_b}(m_a \frac{h}{a} - m_b \frac{h}{b}) \\ &= \frac{s}{m_a + m_b}(m_b \frac{a}{b}\frac{h}{a} - m_b \frac{h}{b}) = 0 \end{aligned} \quad (6)$$

Can these two masses move spontaneously? No. Let us read the Torricelli's explanation. "Non habeant si possibile est aequale momentum, sed altero preponderante moveantur, & ascendat grave *a* versus *c*, descendatque grave *b*. Assumpto iam quolibet puncto *e*, cum grave *a* fuerit in *e*, & *b* in *d*, erunt lineae *ae*, *bd*. aequales, quia idem funiculus est, tam *acb* quam *ecd*. Demptoque communi *ecb* remanent aequales *ae,bd*. Ducatur *ef* parallela ipsi *cd*, & connectantur puncta *e d*. Est igitur grave *a*. ad grave *b*. ut *ac* ad *cb*, hoc est ut *ae* ad *ef*, hoc est *bd*. ad *ef*, hoc est *dg* ad *eg* reciprocè. Est ergo punctum *g* centrum gravitatis commune gravium connexorum, & est in eadem linea horizontali in qua fuerat entequam gravia moverentur. Duo ergo gravia simul colligata mota sunt, & eorum commune centrum gravitatis non discendi. Quod est contra premissam aequilibrii legem."
Torricelli's law of equilibrium of momenta is our equilibrium of forces.

## 6. Momenta and lengths of inclined planes

"Momenta gravium equalium super planis inaequaliter inclinatis, eandem tamen elevationem habentibus, sunt in reciproca ratione cum longitudinibus planorum.

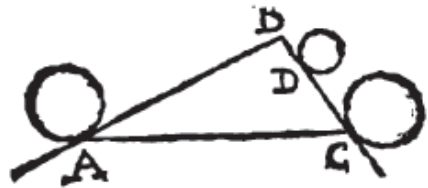

**Fig.5 – Inclined planes and masses**

Sint plana *ab,bc*, inaequaliter inclinata, & ad idem punctum *b*. elevata. Sintque in eisdem planis aequalia gravia *a* & *c* (Fig.5). Dico momentum gravis *c* ad momentum gravis *a* esse reciprocè ut *ab*, ad *bc*. Fiat ut *ab*, ad *bc*, ita grave *a* ad grave aliud *d*, & ponatur *d*. in plano *bc*. Ergò per praecedendetm erunt ipsorum *a*, & *d*. momenta aequalia.
Momentum autem *c*. ad momentum *d*. est ut moles ad molem (quia sunt in eodem plano) hoc est ut moles *a*, ad molem *d*: hoc est ut *ab*, ad *bc*. Est ergo momentum *c*. ad *d*, vel ad momenta *a*. ipsi momento *d*. aequale ut *ab*, ad *bc*. Quod erat &c."
Torricelli is telling that if we consider two equal masses, the two momenta are inversely proportional to the lengths of the planes.

Let us consider the component of the weight parallel to the inclined plane. We have the abovementioned ratio:

$$\frac{m_a g \sin \alpha \cdot \Delta t}{m_c g \sin \beta \cdot \Delta t} = \frac{h}{ab} \frac{bc}{h} = \frac{bc}{ab} \qquad (7)$$

In the case that we have another mass $d$, and $m_a / m_d = ab/bc$:

$$\frac{m_a g \sin \alpha}{m_d g \sin \beta} = \frac{m_a}{m_d} \frac{h}{ab} \frac{bc}{h}$$
$$= \frac{m_a}{m_d} \frac{bc}{ab} = \frac{ab}{bc} \frac{bc}{ab} = 1 \qquad (8)$$

## 7. On times

"Tempora lationum ex quiete per plana eandem elevationem habentia, sunt homologè ut longitudines planorum. Sint plana *ab*,*ac*. eandem elevationem *ad* habentia (Fig.6).

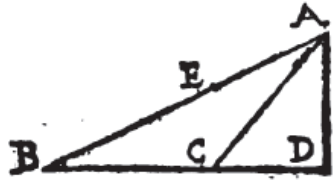

**Fig.6 – Geometry of the problem**

Dico tempus lationis per *ac* ad tempus per *ab* esse ut *ac* ad *ab*." In the case of a point-like mass, neglecting friction, starting from rest (ex quiete), we have in fact that:

$$l_{ab} = \frac{1}{2} a_{ab} t^2 = \frac{g}{2} \sin \alpha \cdot t^2 = \frac{g}{2} \frac{h}{l_{ab}} t_{ab}^2$$
$$l_{ac} = \frac{1}{2} a_{ac} t^2 = \frac{g}{2} \sin \beta \cdot t^2 = \frac{g}{2} \frac{h}{l_{ac}} t_{ac}^2 \qquad (9)$$

And then:

$$l_{ab}^2 = \frac{h}{2} g \cdot t_{ab}^2$$
$$l_{ac}^2 = \frac{h}{2} g \cdot t_{ac}^2 \qquad (10)$$

And therefore: $l_{ab}/l_{ac} = t_{ab}/t_{ac}$, as told by Torricelli.

To solve this problem, Torricelli is using some supposed facts, but, after, he tells that it is possible to use a theorem demonstrated by Galileo:

"Precedens theorema poterat demonstrari sine ulla suppositione. Demonstrat enim Galileus in Prop. 6 de motu accelerato, tempora lationum per chordas omnes in circulo equalia esse." Let us try to investigate what Galileo is proposing.

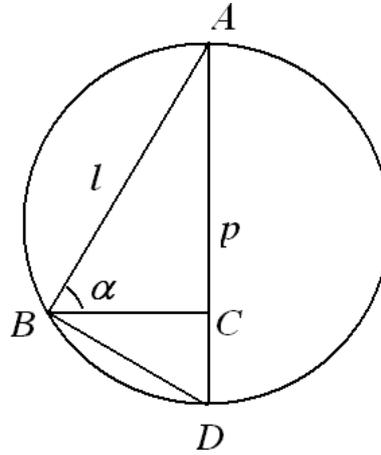

**Fig.7 – Geometry of the problem**

Let us consider the two similar triangles ABC and ABD (Fig.7). AD being the diameter, $p =$ AC, we have: $l/p = 2r/l$. If the particle is starting from rest:

$$2r = \frac{1}{2}g \cdot t_o^2$$
$$l = \frac{1}{2}g \sin\alpha \cdot t_1^2 = \frac{g}{2}\frac{p}{l}t_1^2 \qquad (11)$$

$$2r/l = \frac{l}{p}t_o^2/t_1^2 = \frac{2r}{l}t_o^2/t_1^2$$
$$\rightarrow 1 = t_o^2/t_1^2 \rightarrow t_o^2 = t_1^2 \qquad (12)$$

That is: "tempora lationum per chordas omnes in circulo equalia esse", the time required to descent along any chord is the same.

### 8. On velocity

"Gradus velocitatis eiusdem mobilis super diversas planorum inclinations acquisiti, tunc aequales sunt cum eorundem planorum elevations aequalise sint."

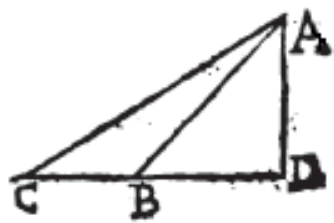

**Fig.8 – Geometry of the problem.**

Let us consider the "gradus velocitatis" as the "modulus of velocity", or "speed". Torricelli is telling that, starting from rest, the speed at the end of the two inclined planes is the same, since the height of the two planes is the same.

We can here remark that Torricelli distinguished kinematics from dynamics, using terms such as acceleration, velocity or impetus in kinematics and momentum in dynamics, when we observe the effects of momentum in the impact forces or in the equilibrium problems.

## 9. The slope of a parabola

"Linae, quae intra parabolam basi parallelae ducuntur (Fig.9), sunt in subdupla ratione portionum diametri ad verticem parabole interceptum."

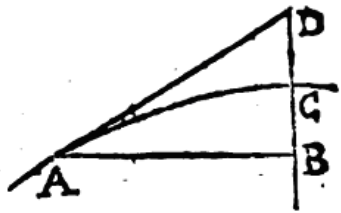

**Fig.9 –A parabola AC.**

Let us add a frame of reference (Fig.10) to the image given in Torricelli's book [9].

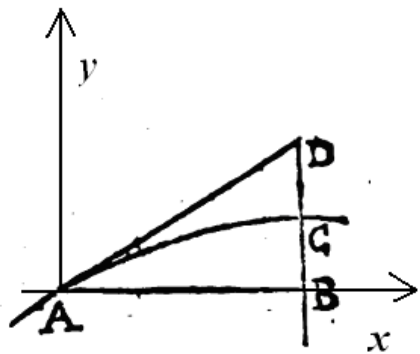

**Fig.10 – Frame of reference for the problem.**

Then we can write the equation of a parabola and its derivative as:

$$y = bx - ax^2$$
$$dy/dx = b - 2ax \quad (13)$$

Moreover:

$$dy/dx = 0 = b - 2ax_v$$
$$dy/dx = b, \quad if \ x = 0 \quad (14)$$

In (14), $x_v = x_C = b/2a$ is the vertex abscissa.

The straight line is $y = bx$ and therefore:

$$y_D = bx_v = b\frac{b}{2a} = \frac{b^2}{2a} \qquad (15)$$

The parabola has the coordinates of its vertex given by:

$$y_C = b\frac{b}{2a} - a\frac{b^2}{4a^2} = \frac{b^2}{4a} = \frac{1}{2}y_D \qquad (16)$$

The result given in (16) is that told by Torricelli. He continued telling also that "Si in parabola aliquod punctum *a* sumatur ex quo linea ducatur basi parallela *ab*, & portioni diametric *bc*. ad verticem intercepte, aequalis recta linea *cd*. ponatur in directum. Recta linea *da*, quae ab extremo positae linae termino *d*: ad punctum *a* in parabola sumptum, ducitur, parabolam continget."

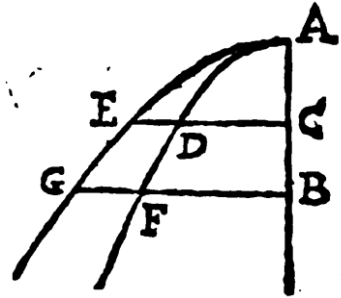

**Fig.11 – On parabolas.**

### 10. On parabola again

"Sunt enim quadrata, *ec*. ad *gb*, ut recta *ac*. ad *ab*. Quadrata etiam *dc* ad *fb* sunt ut *ac*, ad *ab*. Ergo in eadem ratione sunt quadrata inter se; quare ut recta *ec* ad *gb*, ita est *dc* ad *fb*."

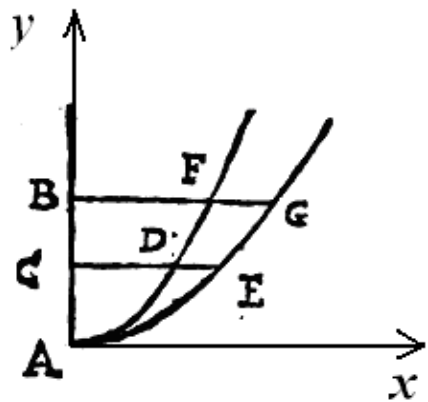

**Fig.12 – A frame of reference added to Fig.11.**

Let us consider the frame *x,y* as in Fig.12, and:

$$y_C = y_D = y_E$$
$$y_B = y_F = y_G \qquad (17)$$

We can use two parabolas: $y = ax^2; y = bx^2$. Then:

$$y_C = ax_D^2; y_C = bx_E^2$$
$$y_B = ax_F^2; y_B = bx_G^2 \qquad (18)$$

$$y_C / y_B = ax_D^2 / ax_F^2 = bx_E^2 / bx_G^2$$
$$\rightarrow x_D^2 / x_F^2 = x_E^2 / x_G^2 \qquad (19)$$

Then we have $x_D / x_F = x_E / x_G$, and therefore, in Torricelli's notation: $dc / fb = ec / gb$.

## 11. On vertical motion

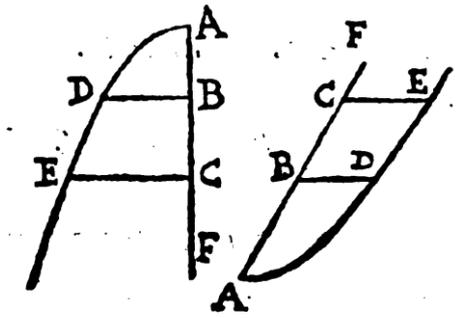

**Fig.13 – Geometry for the study of motion.**

"Sint spatia quaeliter *ab,ac*, sive perpendicularia sive inclinata, & circa diametrum *ac* fiat parabola quaeliter *ade*. atque ordinatim ducantur *bd. ce* (Fig.13). Dico tempus per *ab* ad tempus per *ac* esse ut *bd*, ad *ec*. Sunt enim tempora in subdupla ratione spatiorum ex Galileo, sed linae *db, ec* sunt in subdupla ratione spatiorum (quia quadrata earum sunt ut *ab*, ad *ac*.) ergo eadem ratio est & temporum,& linearum ordinatim ad spatial applicatarum." "Subdupla" means "subduplicata", that is, under the square root [10].

Let us suppose an accelerated motion:

$$\frac{1}{2}at_{AB}^2 = AB; \frac{1}{2}at_{AC}^2 = AC \qquad (20)$$

Then:

$$t_{AB} = \sqrt{\frac{2 \cdot AB}{a}}; t_{AC} = \sqrt{\frac{2 \cdot AC}{a}} \qquad (21)$$

The ratio of times is then:

$$t_{AB}/t_{AC} = \sqrt{\frac{2 \cdot AB}{a}} / \sqrt{\frac{2 \cdot AC}{a}}$$
$$= \sqrt{\frac{ab}{ac}} = \sqrt{\frac{c \cdot x_{AB}^2}{c \cdot x_{AC}^2}} = \frac{x_{AB}}{x_{AC}} = \frac{bd}{ec} \qquad (22)$$

"Hinc manifestus est impetus gravium in fine portionum diametri parabolae, esse inter se vs linea, quae orinatim applicantur ad extrema ipsarum portionum puncta. Sunt enim ex Galileo impetus ut ipsa tempora, sed ordinatim ducta sunt ut ipsa tempora, ergo impetus sun tut ordinatim ducta ec."

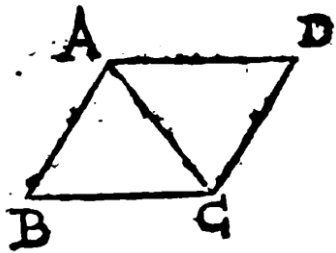

**Fig.14 – Two inclined planes AB and AC.**

### 12. Two inclined planes

"Posito quolibet triangulo *abc*, cuius basis *cb* horizonti parallela sit. Si grave ex quiete in vertice *a* per alterum latus *ac* cadat, & inde per basi a *cd* cum impetus concepto convertatus, basique per acta cum eodem impetus per alterum latum *ba* ascendat, impetus ille perducet grave per ascensum *ba*, usque ad idem punctum *a*. ex quo discesserat." Let us imagine two inclined planes AB and AC (see Fig.14). ABCD is a parallelogram.

Let us assume this time the "impetus" as the kinetic energy. A point-like mass is at rest in A. Let it free to move: it arrives in C and then moves on the plane CB, reaching D. Why? Torricelli tells that the impetus in C is the same of the impetus in B, that the particle gains falling on the inclined plane AB. Then the particle is able arrive at position D because it is the same to revert the motion from A to B in a motion from B to A. We can find such an example in the Lectures to the Accademia too.

### 13. Conclusion
In this paper we have shown, using some parts of the De Motu Gravium Naturaliter Descendentium, how Torricelli could have taught the physics concerning inclined planes and equilibrium. He used only proportions; in some cases, I included trigonometric functions and derivatives to have a more simple comparison. After adjusting the 17th century format to our language and notations, we can easily see that some of the Torricelli's problems are similar to the problems that we are discussing today with our students.

### 14. Appendix
This appendix is reporting what John Gorton, in his general biographical dictionary, is writing about Evangelista Torricelli [2]. Illustrious mathematician and philosopher. Torricelli was born in Faenza, in Italy, October 15, 1608. "He was instructed in Greek and Lain by his uncle, who was a monk, probably with a view to his obtaining preferment in the church; but his genious induced him to devote himself to the study of mathematics, which he attended to for some time without a master; but at the age of twenty he went to Rome, and prosecuted his studies under father Benedict Castelli. Torricelli thus assisted made great improvement, and having read Galileo's Dialogues, he composed a treatise concerning Motion, according to his principles. Castelli, astonished at the ability displayed in this piece, took it to Galileo in Florence, who conceived a high opinion of the author, and engaged him as his amanuensis." John Gorton tells us that Torricelli entered on this office in October 1641, but Galileo dyed three months after. Gorton continues explaining that "Torricelli was about to return to Rome, when the grand duke of Tuscany, Ferdinand II, engaged him to continue at Florence, giving him the title of ducal mathematician, and the promise of a professorship in the university on the first vacancy. Here he applied himself closely to study, and made many improvements and some discoveries in mathematics, physics, and astronomy. He vastly improved the construction of microscopes and telescopes; and he is generally considered as having first ascertained the gravity of the air, by means of mercury in a glass tube, whence resulted the barometer. He would probably have added more to the stores of science if he had not been cut off prematurely, after a few days' illness, in Oct. 1647."